# Epitaxial rare-earth doped complex oxide thin films for infrared applications

*Mythili Surendran[1,2], Joshua Rollag[3,4], Christopher E. Stevens[3,4], Ching-Tai Fu[5], Harish Kumarasubramanian[1], Zhe Wang[6], Darrell G. Schlom[6,7], Ricky Gibson[4], Joshua R. Hendrickson[4] and Jayakanth Ravichandran [1,2,5]\**

**AUTHOR ADDRESS**

[1]Mork Family Department of Chemical Engineering and Materials Science, University of Southern California, 925 Bloom Walk, Los Angeles, CA 90089, USA

[2]Core Center of Excellence in Nano Imaging, University of Southern California, 1002 West Childs Way, MCB Building, Los Angeles, CA 90089, USA

[3]KBR Inc., 3725 Pentagon Blvd, Suite 210, Beavercreek, OH 45431, USA

[4]Air Force Research Laboratory, Sensors Directorate, 2241 Avionics Circle, Wright-Patterson Air Force Base, OH 45433, USA

[5]Ming Hsieh Department of Electrical and Computer Engineering, University of Southern California, 925 Bloom Walk, Los Angeles, CA 90089, USA

[6]Department of Materials Science and Engineering, Cornell University, Ithaca, New York 14853, USA

[7]Kavli Institute at Cornell for Nanoscale Science, Ithaca, New York 14853, USA

**AUTHOR INFORMATION**

**Corresponding Author**

\* E-mail: jayakanr@usc.edu

**ABSTRACT**

Rare earth dopants are one of the most extensively studied optical emission centers for a broad range of applications such as laser optoelectronics, sensing, lighting, and quantum information technologies due to their narrow optical linewidth and exceptional coherence properties. Epitaxial doped oxide thin films can serve as a promising and controlled host to investigate rare-earth dopants suitable for scalable quantum memories, on-chip lasers and amplifiers. Here, we report high-quality epitaxial thin films of Tm-doped $CaZrO_3$ grown by pulsed laser deposition for infrared optoelectronic and quantum memory applications. We perform extensive structural and chemical characterization to probe the crystallinity of the films and the doping behavior. Low temperature photoluminescence measurements show sharp radiative transitions in the short-wave infrared range of 1.75 – 2 μm.

**INTRODUCTION**

Rare-earth ions (REIs) doped in dielectric crystals are extensively studied with applications in laser optoelectronics, solid-state lighting, sensing, displays and most recently, in quantum information[1-9]. The inner 4*f* levels of rare earth ions are shielded from the local environment by the outer full 5*s* and 5*p* orbitals. This results in narrow 4*f*-4*f* transitions that span the far infrared to ultraviolet range. Even when doped into a crystalline host, the efficient shielding of 4f levels ensures that the crystal acts as a very weak perturbation to the free rare-earth ion levels. The electrons in the valence band of REI doped crystals can be efficiently excited to the 4*f* levels and exhibit emission wavelengths longer than near infrared, especially in the mid-IR, making them suitable for various applications such as on-chip lasing. They often have long optical and spin coherence times at cryogenic temperatures, which is essential for quantum information storage. This property helps in maintaining the quantum states for extended periods, allowing for more robust quantum information processing and computing. REIs possess numerous transitions from 1.4 – 12 μm that can be exploited for fiber lasers and amplifiers[3]. Host selection is also extremely important as impurities, dislocations and other lattice imperfections can cause inhomogeneous broadening to these narrow transitions, while phonons and nuclear and electron spins of the host can make significant host-lattice dependent contributions to the homogeneous broadening[1, 5]. Nuclear magnetism contribution to homogeneous broadening can be minimized by using REIs with isotopic abundance close to 100% and hosts comprising of elements with low or zero nuclear magnetic moment or very low isotopic abundance of the magnetic nuclei[1].

Among stable REIs, $Pr^{3+}$, $Tb^{3+}$, $Ho^{3+}$ and $Tm^{3+}$ have 100% isotopic abundance and have strong transitions in the short-wave and mid-wave infrared region which are promising for quantum information storage and on-chip infrared lasers. Epitaxial thin films provide a controlled environment for studying and utilizing the unique properties of REIs and are essential for implementing scalable devices integrated on existing silicon-based platforms. High-quality epitaxial $Y_2O_3$ and $CeO_2$ films have been demonstrated as a promising host for REIs with long coherence times for quantum memory applications[10, 11]. However, the possibility of using REI doped complex or perovskite oxides for quantum and infrared lasing remains unexplored. Perovskite oxides provide a versatile platform for incorporating rare-earth dopants, with large chemical tunability and wide band gaps. The incorporation of REIs introduces well-defined energy levels and transitions that contribute to mid-IR photoluminescence. Additionally, perovskite oxides have structural tolerance to allow for a higher solubility limit for dopants without phase separation, compared to conventional semiconductors, and thus can achieve high dopant levels without clustering or concentration quenching. Moreover, these doped complex oxides can be epitaxially grown using techniques such as molecular beam epitaxy (MBE) and pulsed laser deposition (PLD). The objective of this work is to demonstrate efficient REI doping in epitaxially grown single crystal perovskite oxides and their potential to be used in quantum storage and infrared photonic applications.

Here, we chose $Tm^{3+}$ among the 100% isotopically abundant REIs. The accessible $Tm^{3+}$ emissions in the 2μm (~1.8 – 2μm for $^3F_4 \rightarrow ^3H_6$) spectral region are of interest for numerous applications, such as remote sensing, military applications and medical procedures. $Tm^{3+}$-doped materials, such as crystals and glasses, have been employed in high power and highly tunable laser systems that operate in this mid-IR range[12] and can be easily pumped by ~ 800 nm diode lasers. For optical transitions in REIs, the spectral line shapes are dominated by Stark splitting of the energy levels, which describes the shifting and splitting of energy levels due to the presence of an external electric field[13]. In the case of REIs embedded in a crystalline host, Stark level splitting occurs due to the crystal field of the host resulting in several Stark levels depending on the total angular momentum and number of electrons in the REI. Additionally, local crystal field variations within the host due to multiple crystal site occupancies or defects cause different Stark splitting and results in a broadening of the discrete energy levels. Both $^3F_4$ and $^3H_6$ levels in $Tm^{3+}$ undergoes Stark level splitting resulting in a broad $^3F_4 \rightarrow ^3H_6$ transition with multiple peaks. However, the $^3F_4$ level in $Tm^{3+}$ has a lifetime of about 15 ms at 78 K when doped in YLF single crystals[14], and these long coherence times of $Tm^{3+}$ ions can make them suitable for storing and retrieving quantum information.

The perovskite oxide host chosen here is $CaZrO_3$ (CZO) as Ca and Zr both have high isotopic abundance and low or zero nuclear moment for all isotopes. CZO also has a low phonon energy[15] which is essential to minimize nonradiative transitions through phonon assisted processes. Moreover, CZO has good lattice match with scandate substrates such as $DyScO_3$ (DSO) and $GdScO_3$ (GSO) which also possess low phonon energies (< 500 $cm^{-1}$ in the case of DSO[16]). Epitaxial growth of CZO allows for the precise control of Tm-dopant concentrations in the thin film, enabling the tuning of material properties for specific applications. It also minimizes lattice mismatches and defects, reducing phonon interactions that can lead to non-radiative processes and thus, achieve longer photoluminescence lifetimes and improved overall efficiency. $Tm^{3+}$ ions were doped into the Ca site of CZO. 0.5% and 2% Tm-doped CZO (Tm-CZO) and dense PLD targets were synthesized by solid-state reaction. Tm-CZO thin films were also grown on $SrTiO_3$ (STO)/Si substrates for potential silicon photonics integration. These epitaxial thin films can be integrated with semiconductor devices and quantum circuits, enabling the development of on-chip mid-IR light sources and detectors, as well as quantum computing applications.

**EXPERIMENTAL METHODS**

**Thin film deposition.** The thin films were grown by PLD using a 248 nm KrF excimer laser in a vacuum chamber. Single crystal (Crystec GmbH) $DyScO_3$ (110) substrates were pretreated by annealing at 1000°C for 3 hours in an oxygen environment with a flow rate of 100 sccm and subsequently cleaned in acetone and IPA prior to deposition. The chamber was evacuated to a base pressure of ~ $10^{-7}$ mbar. The substrate was heated up to the growth temperature in an oxygen partial pressure of 5x10$^{-3}$ mbar. High purity stoichiometric powders of CaCO$_3$ (99.997%, Alfa Aesar), ZrO$_2$ (99.978%, Alfa Aesar) and Tm$_2$O$_3$ (99.99%, Alfa Aesar) were ground in a mortar and pestle and calcined at 1200°C for 10h to form 0.5% and 2% Tm-doped CaZrO$_3$ (Tm-CZO) powders. In the second step, stoichiometric Tm-CZO powder was pressed into ¾ inch pellet and densified to > 95% density by sintering at 1500°C for 40h. These dense pellets were used as the targets and were pre-ablated before growth. We used *in situ* reflection high energy electron diffraction (RHEED) to monitor the sample surface during the growth as well as to obtain growth rate and film thickness. The fluence was fixed at 1.5 J/cm$^2$ and the target substrate distance used was 75 mm. The films were cooled postgrowth at a rate of 5°C/min at an oxygen partial pressure of 100 mbar.

**Structural and surface characterization.** The high resolution out-of-plane X-ray diffraction (XRD) and reciprocal space mapping (RSM) were carried out on a Bruker D8 Advance diffractometer using a Ge (004) two bounce monochromator with Cu K$\alpha$1($\lambda$ = 1.5406 Å) radiation at room temperature. X-ray reflectivity (XRR) measurements were done on the same diffractometer in a parallel beam geometry using a Göbel (parabolic) mirror set up. STEM sample preparation was performed using a Thermo Scientific Helios 5 EXL Dual Beam equipped with an EasyLift manipulator. Standard *in situ* lift-out technique was used to prepare the TEM lamella. The lift-out region was first coated with the thin e-beam deposited C and W film and followed by a thicker 1.5 μm ion-beam deposited W coating to avoid ion-beam damage to the film. The sample was milled and thinned down to about 100 nm using a 30 kV beam and final polishing to remove residual surface beam damage was carried out using a 3 kV ion-beam. Scanning tunneling electron microscopy (STEM) experiments were carried out using the probe-corrected Thermo Fisher Scientific Spectra 200 (operated at 200kV) microscope equipped with a fifth-order aberration corrector and a cold field emission electron gun. Energy dispersive X-ray spectroscopy (EDS) was carried out using a Dual-X Bruker EDX detectors.

**Optical Spectroscopy.** The room temperature photoluminescence (PL) spectroscopy measurements were performed in a Renishaw inVia confocal Raman Microscope using a 532 nm diode laser through a 100X objective with a numerical aperture of 0.1. The excitation power used was ~1 mW. Low temperature PL measurements were carried out at 10 K in a cryostat. The pump laser used was 793 nm Ti: Sapphire CW laser with a linewidth of 50 kHz and an output power of 250 mW. The laser passed through a 50/50 beam splitter and a focusing objective that was coated for the visible range, resulting in a transmission of ~40%. This focusing objective had a magnification of 100X and a numerical aperture of 0.75. Thus, the incident power on the samples was ~80 mW and the beam size on the samples was ~1 μm. The PL from the samples passed through the same focusing objective and a 1450 nm long pass filter before entering the spectrometer. The spectrometer used a grating with groove density of 300/mm. The detector used with this spectrometer was a liquid nitrogen cooled, extended InGaAs linear detector. The detector array had 1024 pixels and dimensions of 25 μm by 500 μm.

**RESULTS AND DISCUSSION**

Figure 1a shows a high resolution out-of-plane $2\theta$-$\theta$ scan of 2% Tm-CZO thin film grown on DSO substrate at room temperature (blue solid lines). Only *hh0* type of reflections of CZO were observed indicating strong out-of-plane texture. No impurity peaks from $Tm_2O_3$ were observed confirming that no phase separation occurred during thin film growth. The inset of Figure 1a shows the zoomed in XRD scans of Tm-CZO 110 reflection along with DSO 110 showing Pendellösung fringes corresponding to extremely smooth film surfaces and interface. The rocking curve FWHM for Tm-CZO 110 reflection was 0.03° (figure 1b) indicating highly oriented grains with minimal incoherencies. However, it is worth noting that the lattice mismatch between Tm-CZO and DSO substrate is low (about 1.2%) and the film is expected to be fully strained to the substrate (as shown by the RSM measurements in the next section). However, despite no strain relaxation, the mosaic spread of the grains, density of dislocations, and other defects at the substrate-film interface can contribute to small incoherencies, and consequently, the rocking curve broadening. The XRR curve in figure 1c shows a slow decay of reflected X-ray intensity and presence of extended Kiessig fringes indicating a very low surface and interfacial roughness. Tm-CZO films were also grown on STO/Si substrates and a similar texture was observed as shown in figure 1a where only Tm-CZO $hh0$ peaks can be observed in the out-of-plane direction along with STO 002 and Si 004 reflections (green solid lines).

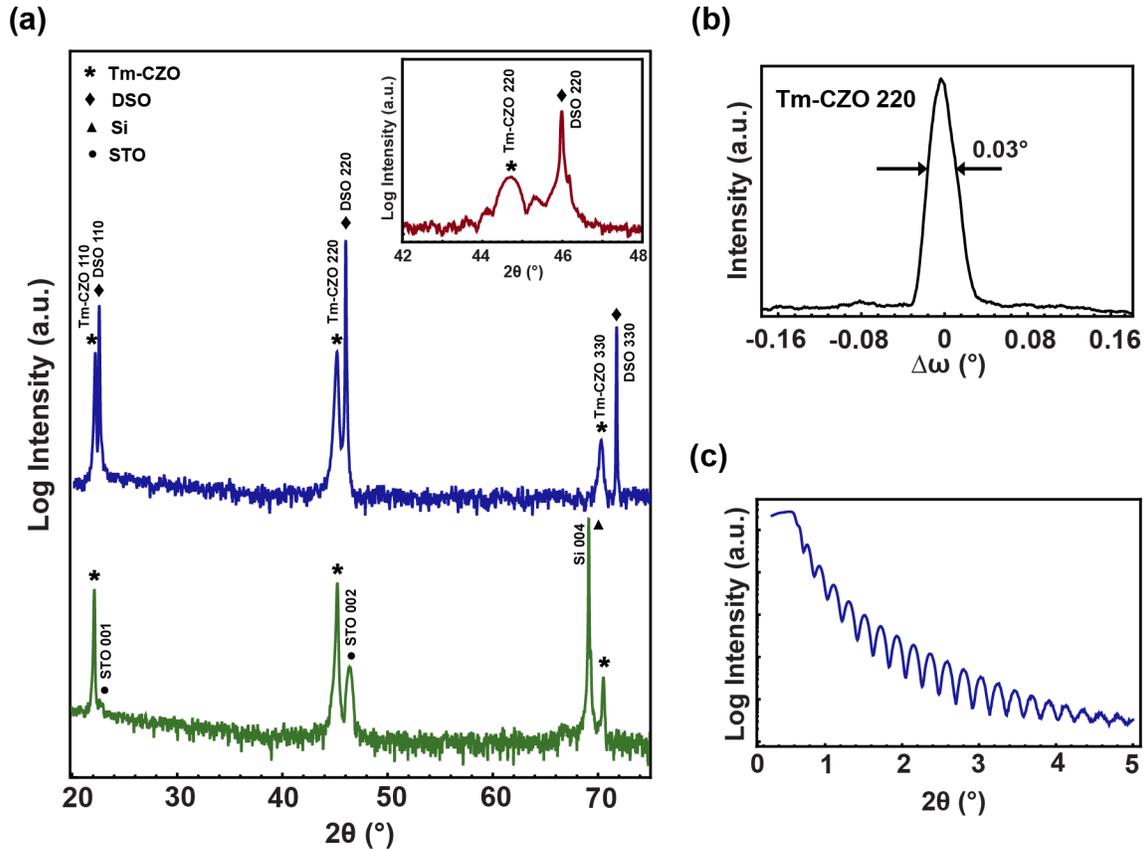

**Figure 1.** (a) High resolution XRD pattern of a representative Tm-CZO film grown on DSO substrate and STO/Si substrate indicating strong out-plane-texture. The inset shows a zoomed in version clearly showing the Pendellösung fringes. (b) Rocking curve indicating a narrow FWHM of 0.03°. (c) XRR curve for a 40 nm Tm-CZO film.

We performed *in situ* RHEED during the Tm-CZO growth to monitor the growth rate and growth mode evolution during PLD. Prior to deposition, annealed DSO substrate showed sharp specular spot 2D diffraction spots (Figure 2a) indicating a smooth, highly oriented substrate surface. During Tm-CZO growth, a well-defined streaky pattern was observed throughout, along with RHEED intensity oscillations. A representative streaky RHEED pattern of a 40 nm Tm-CZO film post growth is shown in Figure 2b indicating a relatively smooth film interface. The corresponding RHEED intensity oscillations of the specular spot are shown in Figure 2c as an evidence for layer-by-layer growth. To determine the epitaxial relationship between the film and the substrate, we performed RSM measurements. The RSM measurements were carried out on a 40 nm Tm-CZO sample and centered around DSO 332 reflection. Both the Tm-CZO 332 and DSO 332 reflections observed in the RSM possessed the same in-plane lattice parameter indicating that the Tm-CZO layer is fully strained to the substrate. The out-of-plane lattice parameter observed in the RSM agreed well with the out-of-plane lattice parameter obtained from the high-resolution $2\theta$-$\theta$ XRD scans. Thus, we conclude that the Tm-CZO films are epitaxially grown on DSO substrate and the epitaxial relationship is Tm-CZO [001] (110) // DSO [001] (110).

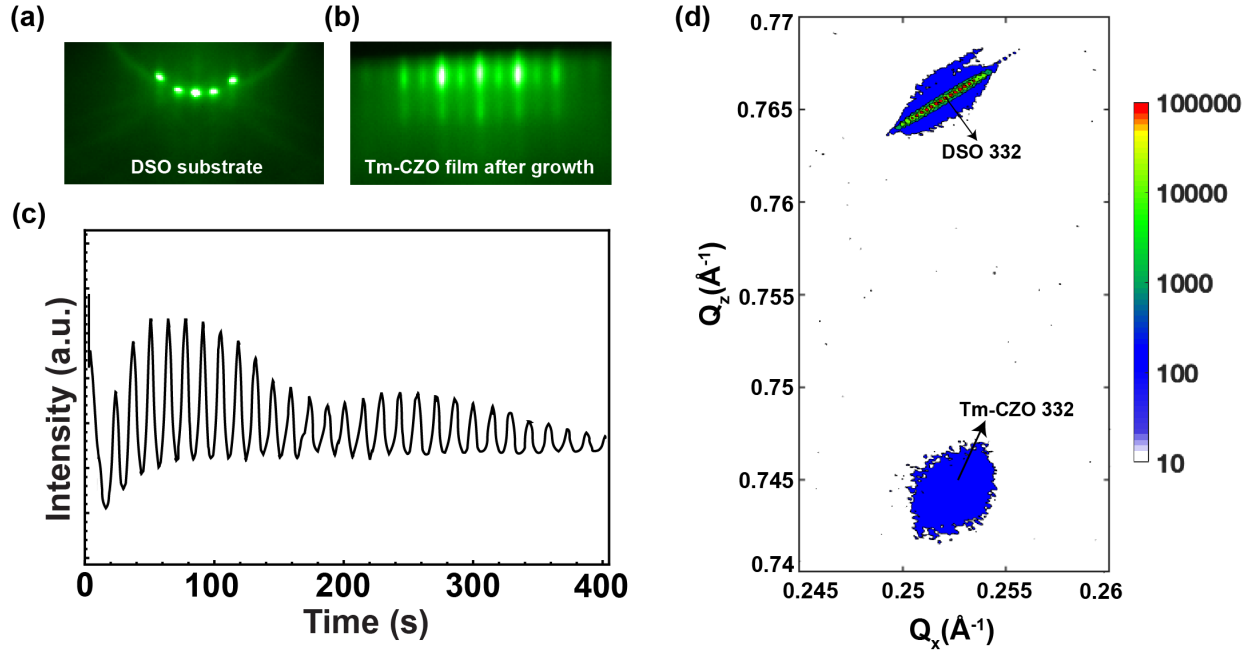

**Figure 2**. a) Representative RHEED pattern for (a) annealed DSO substrate prior to deposition and (b) Tm-CZO thin film after deposition showing streaky pattern. (c) RHEED specular spot intensity oscillations during Tm-CZO growth. (d) A high-resolution reciprocal space map of Tm-CZO thin film centered on DSO 332 substrate peak. The map clearly shows that the film is coherently strained to the substrate and confirms the epitaxial relationship.

STEM experiments were also carried to investigate the atomic structure of Tm-CZO. Figure 3a shows a wide field of view HAADF image of 60 nm 0.5% Tm-CZO film on DSO substrate with a sharp interface. A representative high resolution HAADF image in Figure 3b shows some Ca columns with higher intensity than the neighboring columns (yellow circles) indicating the presence of Tm atoms in these sites. This confirms that the Tm atoms are substituting the A-site Ca atoms as expected. The averaged line profile corresponding to the yellow line in Figure 3b is shown in Figure 3c, showing a higher intensity for the heavier Tm atom site compared to the neighboring lighter Ca sites. Figure 3d shows an atomic resolution HAADF image of Tm-CZO film with EDS maps clearly showing Ca and Zr columns, along with Tm substituted site. The Tm-site with higher intensity indicated by the yellow circle in the HAADF image can be clearly observed in the Tm elemental color map.

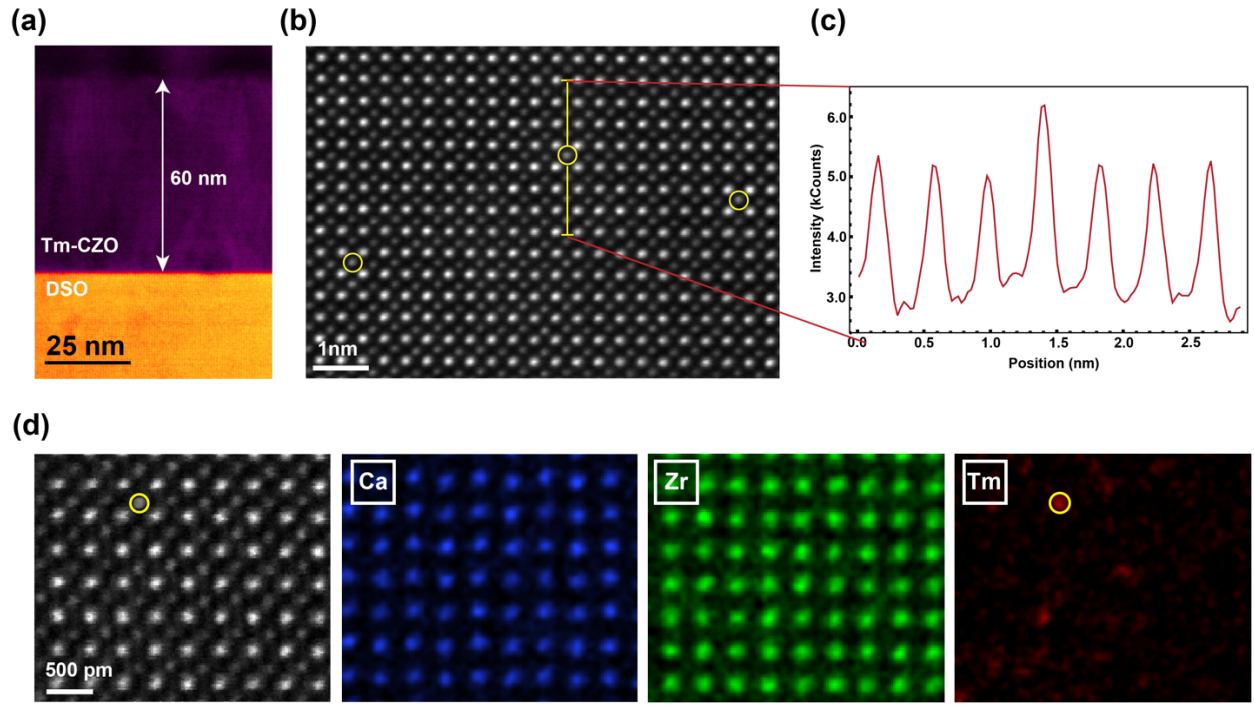

**Figure 3**. (a) Wide field-of-view HAADF image showing 60 nm Tm-CZO film on DSO substrate. (b) Atomic resolution HAADF image of Tm-CZO showing Tm atoms substituting the Ca sites as shown by yellow circles. (c) Intensity line profile along the yellow line in (b) showing a higher intensity for Tm occupied Ca sites. (d) Atomic resolution HAADF image and the corresponding EDS elemental color maps for Ca, Zr and Tm.

For visible/near-infrared PL measurements, 0.5% or 2% doped Tm-CZO on DSO and STO/Si substrates were used. Figure 4a shows the room temperature PL of a 150 nm Tm-CZO film on STO/Si substrate in the visible/near-infrared region. Under excitation of a 532 nm laser, a strong PL peak around 757 nm (1.64 eV) was observed corresponding to $^3H_4 \rightarrow {}^3H_6$ transition of $Tm^{3+}$. A lower intensity PL peak can also be seen around 657 nm (1.88 eV) suggesting the presence of a $^3F_{2,3} \rightarrow {}^3H_6$ radiative transition. However, for low temperature short-wave infrared (SWIR) PL, a 2% Tm-doped CZO thin film (150 nm thick) grown on STO/Si was used in order to increase the signal to noise ratio and avoid overlapping PL peaks from the DSO substrate in the SWIR region. Under an excitation of 793 nm, peaks corresponding to the $^3F_4$ energy level transitions in the range of 1.4 – 1.6 μm and 1.75 – 2 μm were observed. Figure 4b shows PL intensity measured at 10K where the most intense peak was observed at around 1770 nm and corresponds to $^3F_4 \rightarrow {}^3H_6$ transition. The $^3F_4$ and $^3H_6$ manifolds in $Tm^{3+}$ typically undergo Stark level splitting due to the host crystal field and results in a set of peaks in the range 1.75 – 2 μm. The $^3H_4 \rightarrow {}^3F_4$ manifold transitions also appeared as a result of a cross-relaxation mechanism with PL peaks in the range 1.4 – 1.6 μm. Nevertheless, a higher concentration of Tm-doping or thicker film or confining the film in a cavity is needed to measure the excited state lifetime of the $^3F_4$ energy level due to the poor signal to noise ratio in the SWIR region.

These results show that Tm-doped epitaxial CZO films can show strong near to short wave infrared emission with the possibility of MIR emissions, and the strategy can be extended to explore other suitable REIs as well for future applications.

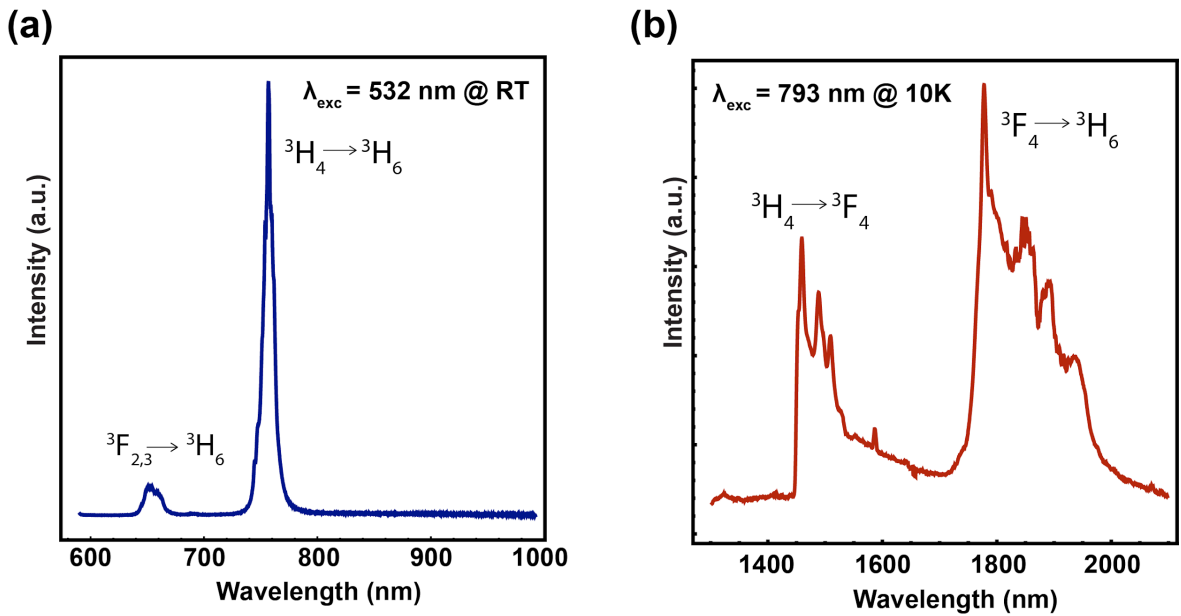

**Figure 4**. (a) Room temperature PL spectrum of 2% Tm-CZO on STO/Si substrate under 532 nm excitation and (b) the corresponding PL spectra at 10 K under 793 nm excitation, showing strong emission in the visible/near-infrared as well as SWIR region.

## CONCLUSIONS

In this work, we demonstrated Tm-doped epitaxial CZO thin films as a potential material system for future mid-infrared on-chip applications. The epitaxial growth ensures high-quality host structure and an efficient doping strategy, while the incorporation of $Tm^{3+}$ introduces properties suitable for laser applications and quantum information processing/storage. Epitaxial Tm-doped thin films were successfully grown on DSO and Si/STO substrates by pulsed laser deposition. The films were fully strained and had a narrow rocking curve as indicated by X-ray diffraction studies. Efficient A-site substitution doping of $Tm^{3+}$ in CZO was confirmed by high-resolution STEM and EDS measurements. The findings suggest that Tm-doped CZO thin films have strong infrared emissions in the 2 μm region and thus, hold promise for technological advancements in the short-wavelength infrared spectrum, contributing to the development of lasers and quantum technologies. Further studies involve lifetime measurements for $Tm^{3+}$ doped CZO thin films, as well as exploring other REIs such as $Er^{3+}$ (emission in the telecom window), $Ho^{3+}$, etc. and optimize the material's properties for specific applications within these fields.

## AUTHOR INFORMATION

**Corresponding Author**

Email: j.ravichandran@usc.edu

**Author Contributions**

M.S. and J. Ravichandran conceived the idea and designed the experiments. M.S and C.T.F synthesized the PLD targets. M.S and H.K performed the thin film growth. M.S performed room temperature PL measurement. J. Rollag, C.E.S, R.G and J.R.H designed and performed the low temperature PL measurements. Z.W and D.S grew epitaxial STO on Si substrates. All authors discussed the results. M.S. and J. Ravichandran wrote the manuscript with input from all other authors.

## ACKNOWLEDGMENT

This work was supported in part by Air Force Office of Scientific Research grant nos. FA9550-23-1-0524 and FA9550-16-1-0335. The authors acknowledge Dr. Anna Carlsson from ThermoFisher Scientific for performing the high-resolution STEM imaging and EDS spectroscopy. The authors gratefully acknowledge the use of facilities at Prof. Stephen Cronin's Lab and Core Center for Excellence in Nano Imaging at University of Southern California for the results reported in this manuscript. J.R.H. acknowledges support from the Air Force Office of Scientific Research (Program Manager Dr. Gernot Pomrenke) under award number FA9550-20RYCOR059.

# Table of Contents Graphic

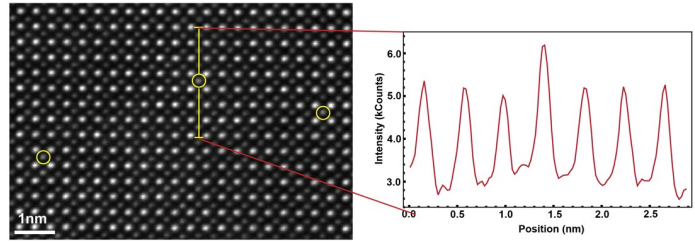
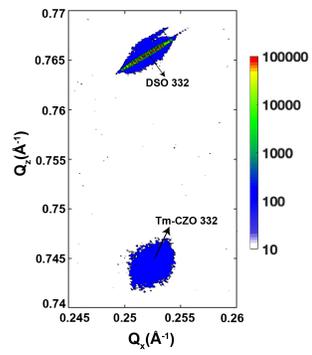
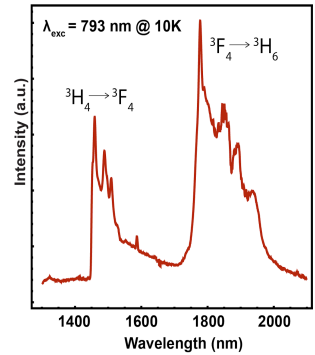